\documentstyle[prl,aps]{revtex}

\begin{document}

\twocolumn[{
\draft
\widetext
\title
{Dynamical description of the buildup process in resonant tunneling: 
evidence of exponential and non-exponential contributions }
\author{Roberto Romo and Jorge Villavicencio}
 
\address{Facultad de Ciencias,
Universidad Aut\'onoma de Baja California\\
Apartado Postal 1880, 22800 Ensenada, Baja California, M\'exico}

\date{30 march 1999}

\mediumtext

\begin{abstract}
The buildup process of the probability density inside the quantum well of a
double-barrier resonant structure is studied by considering the analytic
solution of the time dependent Schr\"{o}dinger equation with the initial
condition of a cutoff plane wave. For one level systems  at resonance condition
we show that the buildup of the probability density
obeys a simple charging up law, $\left| \Psi \left(\tau \right) 
/ \phi  \right| =1-e^{-\tau /\tau _0},$ where $\phi$ is the stationary 
wave function and  the transient time constant $\tau _0$ is exactly two lifetimes. We
illustrate that the above formula holds both for symmetrical and
asymmetrical potential profiles with typical parameters, and even for
incidence at different resonance energies. 
Theoretical evidence of a crossover to non-exponential buildup is also discussed.
\end{abstract}
\maketitle

}]

\narrowtext
Since the pioneering work of Esaki and Tsu\cite{EsakiTsu}, the tunneling in
one-dimensional semiconductor heterostructures has been the subject of
intense investigations \cite{tunneling,capacitor,Tsuchiya,Ricazbel}.
Resonant tunneling in double barrier (DB) systems has received special
attention both by its technological applications and by the motivation to
clarify the new interesting transport phenomena. Among the fundamental
problems that have appeared on the scene, the charge buildup in the quantum
well region is considered as one of the most important processes since it
governs the ultimate speed of resonant tunneling devices \cite
{capacitor,Tsuchiya}. The need of a direct and comprehensive dynamical study
of this phenomenon has been widely recognized \cite{Tsuchiya,Ricazbel,Sakaki}%
. However, up to now we are lacking of an exact description of the buildup
process itself.

In this paper we provide an exact description of the buildup process at
resonance condition, within the framework of the shutter model. This is based 
on a full quantum dynamical approach,
recently developed by Garc\'{\i }a-Calder\'{o}n and Rubio \cite{PRA97}, that
deals with the solution of the time dependent Schr\"{o}dinger equation for
an arbitrary potential $V(x)$ $(0<x<L)$, with an initial condition of a
cutoff plane wave confined in the half-space $x<0$ to the left of an
absorbing shutter \cite{comment} at $x=0$. The sudden opening of the shutter
at $t=0$ allows the wave function to interact with the potential. As a
consequence, they found that the transient solution for the internal region
may be written as the stationary solution modulated by a time varying
Moshinsky function plus an infinite sum of transient resonance terms
associated with the $S-$matrix poles of the problem.

For the case of a reflecting shutter

\begin{equation}
\Psi \left( x,k;t=0\right) =\left\{ 
\begin{array}{cc}
e^{ikx}-e^{-ikx} & \quad -\infty <x\leq 0 \\ 
0 & \quad x>0,
\end{array}
\right.  \label{obtur}
\end{equation}
one can proceed along the lines similar to that discussed in Ref. \cite
{PRA97} and obtain the solution for the internal region,

\begin{eqnarray}
\Psi (x,k;t) &=&\phi (x,k)M(0,k;t)-\phi ^{*}(x,k)M(0,-k;t)  \nonumber \\
&&-i\sum\limits_{n=-\infty }^\infty T_nM(0,k_n;t),  \label{Psiint}
\end{eqnarray}
where $\phi (x,k)$ is the stationary wave function and the factors $%
T_n=2ku_n(0)u_n(x)/(k^2-k_n^2)$ are given in terms of the resonant
eigenfunctions $u_n(x)$. The index $n$ runs over the complex poles $k_n$
distributed in the third and fourth quadrants in the complex $k$-plane. The
Moshinsky functions \cite{PRA97}, as it is well known, are defined in terms
of the complex error function $w(z)$ \cite{Abramwtz}:$\ M(y_q)\equiv
M(0,q;t)=w(iy_q)/2$, where the argument $y_q$ is given by

\begin{equation}
y_q=-e^{-i\pi /4}\left( \frac m{2\hbar }\right) ^{1/2}\left[ \frac{\hbar q}m%
t^{1/2}\right] ,  \label{ye}
\end{equation}
and $q$ stands either for $\pm k$ or $k_{\pm n}$.

The time evolution of wavefunction $\Psi (x,k;t)$ in the internal region may
be described by the expression given by Eq. (\ref{Psiint}), which involves
the contribution of the full resonant spectrum of the system. For structures
with typical parameters, and incidence energies $E=\hbar ^2k^2/2m$ near a
resonance energy $\varepsilon _n$, Garc\'{\i }a-Calder\'{o}n and Rubio \cite
{PRA97} showed that the single resonance approximation for the wavefunction $%
\Psi (x,k;t)$ is valid from a few tenths of the corresponding lifetime
onwards. This is the case for the present study, since we are not
considering the regime of very short times ($t\ll \tau _n=\hbar /\Gamma _n$%
), which may require the contribution of far away resonances. The single
resonance approximation to Eq. (\ref{Psiint}) is 
\begin{eqnarray}
\Psi \left( x,k;t\right) &=&\phi (x,k)M(0,k;t)-\phi ^{*}(x,k)M(0,-k;t) 
\nonumber \\
&&\ -iT_nM(0,k_n;t)-iT_{-n}M(0,-k_n^{*};t),  \label{Psi1term}
\end{eqnarray}
where we have used the fact that the poles in the third quadrant $k_{-n}$
are related to those of the fourth, $k_n$, by $k_{-n}=-k_n^{*}$. This is the
one-level expression of the time dependent wave function for the description
of the dynamics in the internal region.

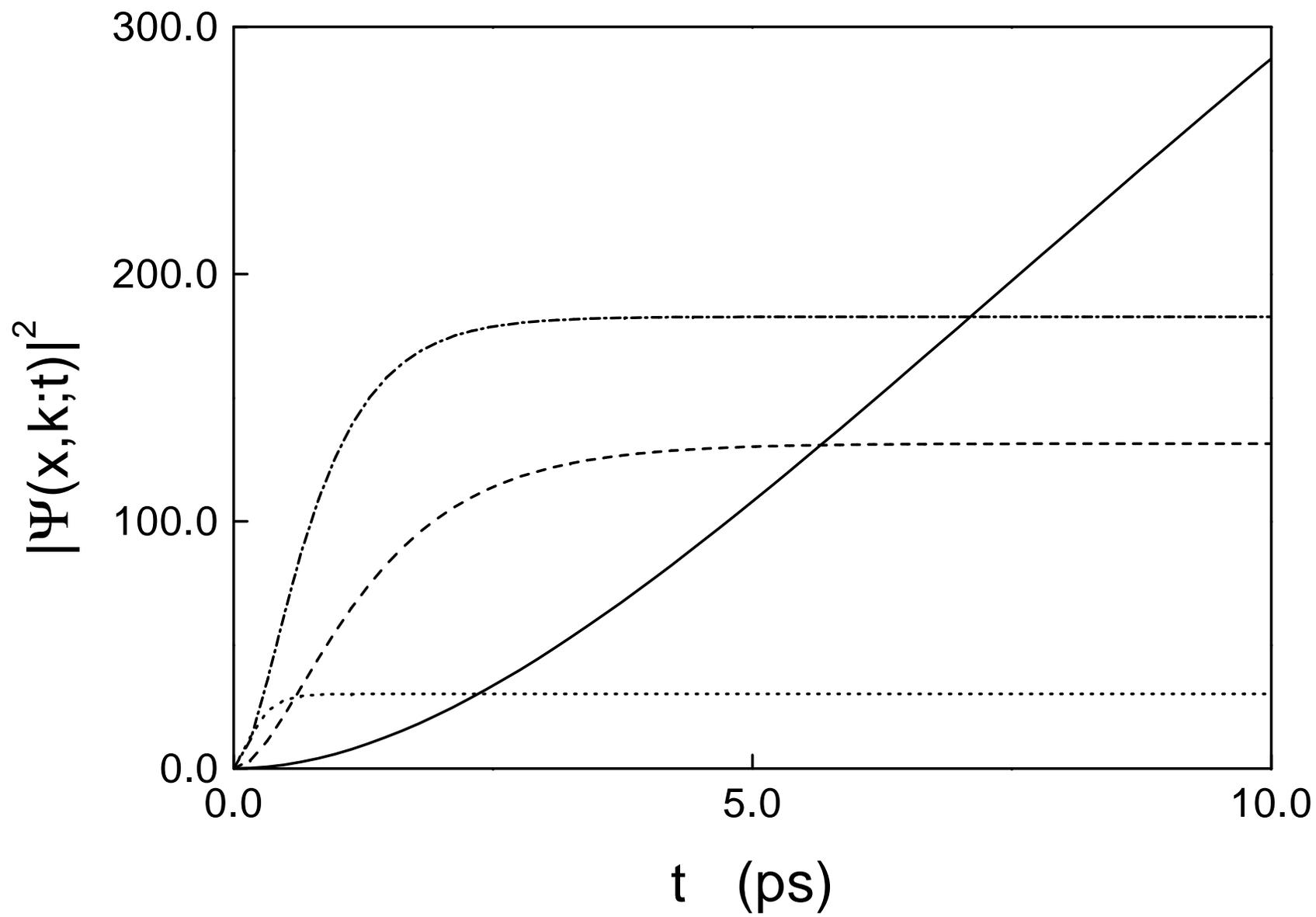
\begin{figure}
\caption{ This graph illustrates the building up of the probability density
for the two examples discussed in the text. In the symmetrical system we
show the time evolution for $|\Psi \left( x,k;t \right) |^2$ for incidence
at: $\varepsilon _1=37.8$ $meV$ (solid line), $\varepsilon _2=149.2 $ $meV$
(dashed line) and $\varepsilon _3=325.7$ $meV$ (dotted line). The fixed
positions $x$ are: $80$ \AA , $48$ \AA\ and $80$ \AA , respectively. In the
asymmetric case $\varepsilon _1=89.1$ $meV$ (dashed-dotted line) at $x=55$
\AA }
\label{fig1}
\end{figure}

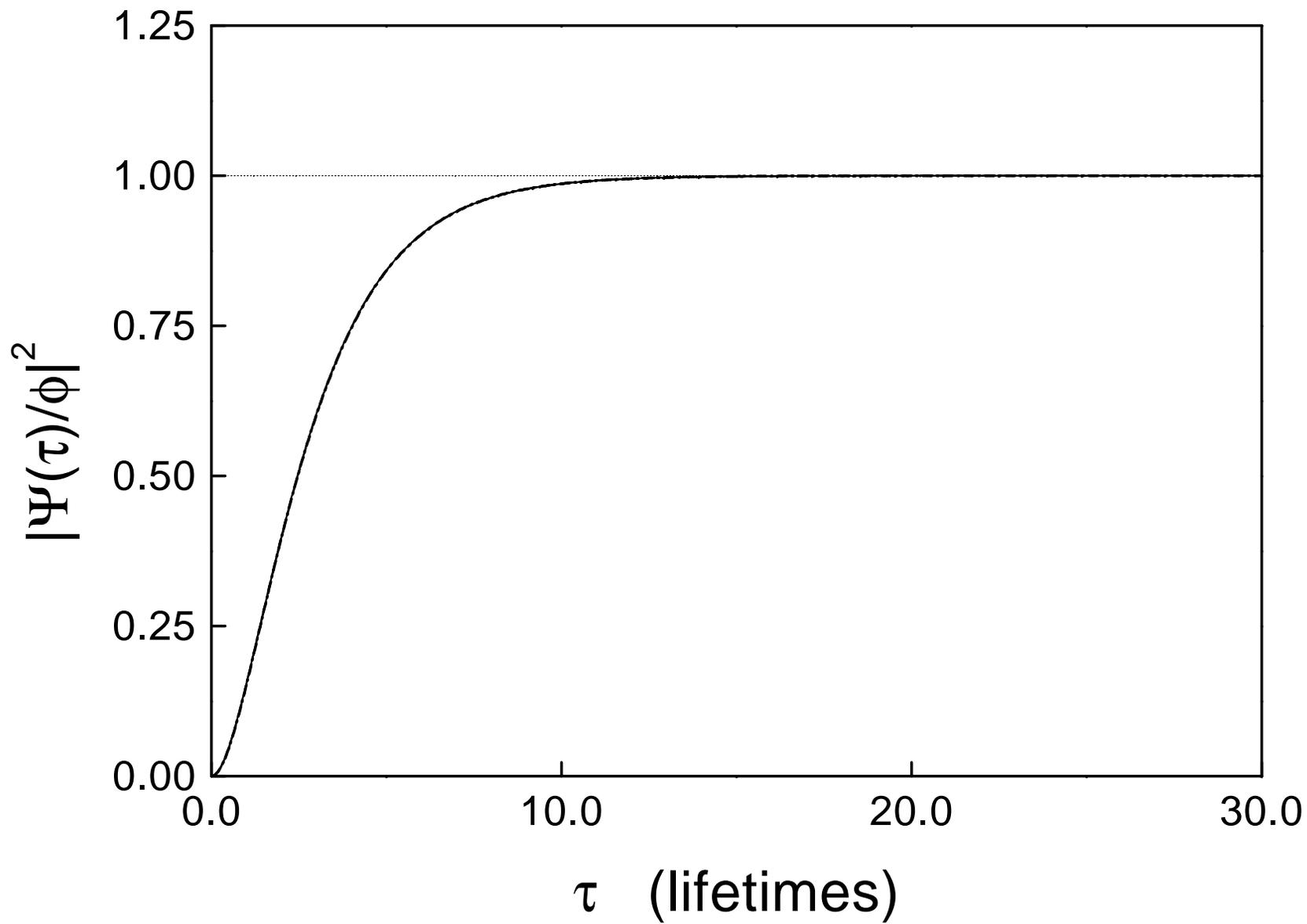
\begin{figure}
\caption{ Evolution of $|\Psi \left( \tau \right) /\phi |^2$ as a function
of $\tau $ for the same cases of Fig.1. All four curves become identical.
For comparison, a plot of $|\Psi \left( \tau \right) /\phi |^2$ using Eq.(6)
is also included. The resulting values are also superimposed on the other
curves illustrating that they obey a simple exponential law with a unique
transient time constant of two lifetimes. }
\label{fig2}
\end{figure}

In order to exemplify the building up of the probability density in the
quantum well for incidence at different resonance energies and for different
potential profiles, let us consider the following two numerical examples.
The first case corresponds to the symmetrical DB structure with parameters:
barrier heights $V_0=0.5$ $eV,$ barrier widths $b_0=30$ \AA\ and well width $%
\omega _0=100$ \AA . The resonance parameters for the first three resonant
states are: $\varepsilon _1=37.8$ $meV$, $\Gamma _1=0.12$ $meV$; $%
\varepsilon _2=149.2$ $meV$, $\Gamma _2=1.40$ $meV$; $\varepsilon _3=325.7$ $%
meV$, $\Gamma _3=8.60$ $meV$.{\bf \ }We show in Fig. 1 the time evolution of
the probability density calculated by Eq. (\ref{Psi1term}) for incidence at
resonance, $E=\varepsilon _n$ and fixed position $x$ (we have considered
values of $x$ near the maxima of $|\phi \left( x,k\right) |^2$ as the most
natural choice) for the cases: $n=1$, $x=80$ \AA\ (solid line); $n=2$, $x=48$
\AA\ (dashed line); and $n=3$, $x=80$ \AA\ (dotted line). The second example
consists of an asymmetrical DB structure with parameters: barrier heights $%
V_1=V_2=0.3$ $eV,$ barrier widths $b_1=30$ \AA\ and $b_2=100$ \AA , and well
width $\omega _0=50$ \AA . The resonance parameters for the first resonant
state are $\varepsilon _1=89.1$ $meV$ and $\Gamma _1=2.4$ $meV$.{\bf \ }The
time evolution of $|\Psi \left( x,k;t\right) |^2$ for incidence at $%
E=\varepsilon _1$ and $x=55$ \AA\ is also depicted in Fig. \ref{fig1}
(dashed-dotted line). For all cases, the probability density $|\Psi \left(
x,k;t\right) |^2$ grows up monotonically towards its asymptotic value. We
can see that both the level off and the rate of increase of the curves are
quite different. However, a common feature, not evident in Fig. \ref{fig1},
can be appreciated if we replot the normalized probability density $|\Psi
\left( x,k;\tau \right) /\phi (x,k)|^2$ as a function of the new variable $%
\tau $, which is now the time given in lifetime units ($t$ replaced by $\tau
\hbar /\Gamma _n$ for each curve). As a result, all four curves become
indistinguishable among them as depicted in Fig. \ref{fig2}. We can see that
the full establishment of the stationary situation is preceded by a
transient in which the probability density is built up inside the quantum
well with a unique characteristic curve. Thus, there must also exist a
characteristic transient time constant $\tau _0$ that governs the buildup
process, with the same value (in lifetimes) for all cases. The observed
regularity and the fact that it holds for both symmetrical and asymmetrical
cases and at different resonances, is a manifestation that the buildup
process in one-level systems is governed by a simple law. In what follows we
shall be concerned to find the analytic expression of the actual buildup law
and the exact value of the characteristic transient time $\tau _0$.

For sharp ($R_n\equiv \varepsilon _n/\Gamma _n\ \gg 1$) and isolated
resonances we know that the stationary wavefunction $\phi (x,k)$ can be
written as the one-term expression \cite{GGC87} $\phi (x,k)=2iku_n\left(
0\right) u_n\left( x\right) /(k^2-k_n^2)$, thus the factors $iT_n$ and $%
iT_{-n}$ appearing in Eq. (\ref{Psi1term}) can be identified as $\phi (x,k)$
and $-\phi ^{*}(x,k)$ respectively. By expressing formula (\ref{ye}) in
lifetime units, it can be seen that $y_q$ depends only on the ratio $%
R_n=\varepsilon _n/\Gamma _n$, and not on the particular values of the
resonance parameters $\varepsilon _n$ and $\Gamma _n$. For $q=\pm k_n$ and $%
q=\pm k_n^{*}$, $y_q$ reads: $y_{\pm k_n}=\mp e^{-i\pi /4}\left[ \left(
R_n-i/2\right) \tau \right] ^{1/2}$, and $y_{\pm k_n^{*}}=\mp e^{-i\pi
/4}\left[ \left( R_n+i/2\right) \tau \right] ^{1/2}$, respectively. For the
cases $q=\pm k$ we have $y_{\pm k}=\mp e^{-i\pi /4}\left[ R_n\tau \right]
^{1/2}$ (since $E=\varepsilon _n$). From the above considerations and the
well known symmetry relation \cite{PRA97}, $M(y_q)=e^{y_q^2}-M(-y_q),$
applied to the Moshinsky functions $M(y_k)$ and $M(y_{k_n})$, we obtain a
convenient representation for the probability density, 
\begin{equation}
\left| \Psi \left( x,k;\tau \right) \right| ^2=\left| \phi (x,k)\right|
^2(1-e^{-\tau /2})^2+\Delta (\tau ),  \label{Psiap2}
\end{equation}
where $\Delta (\tau )$ stands for the remaining terms, which involve the
square modulus of the Moshinsky functions and several interference terms. It
is not difficult to convince oneself that, for very large times, expression (%
\ref{Psiap2}) possesses the correct asymptotic behavior, {\it i. e.} as $%
\tau \rightarrow \infty ,${\it \ }$|\Psi \left( x,k;\tau \right) |^2$ goes
into the stationary probability density $|\phi (x,k)|^2$. This follows
directly from the presence of the decreasing exponential $e^{-\tau /2},$ and
from the fact that each of the Moshinsky functions involved in $\Delta (\tau
)$ can be represented by a series expansion consisting of inverse powers of $%
\tau $. In Ref. \cite{PRA97} it was shown that $M(y_q)$ has the asymptotic
expansion $M(y_q)=a_1/y_q+a_2/y_q^2+a_3/y_q^3+\cdot \cdot \cdot $ for large
values of the variable $y_q$ provided that $-\pi /2<arg(y_q)<\pi /2$. By
inspection of our expressions for $y_q$, we can see that the above
inequality holds for the three cases involved in $\Delta (\tau )$ $(q=-k,$ $%
-k_n$, and $-k_n^{*})$, and as a consequence $\Delta (\tau )\rightarrow 0$
when $\tau \rightarrow \infty $, as expected. It is important not only that
both $e^{-\tau /2}$ and $\Delta (\tau )$ go to zero, but also the fact that
their rates of decrease are quite different, leading to important
consequences on the nature of the buildup. In particular it reveals that
there exist exponential and non-exponential contributions to the buildup
mechanism, as we shall see later. In view of the series expansions
considered above, $\Delta (\tau )$ also contains inverse powers of $y_q,$
that is, inverse powers of the product $(R_n\tau )^{1/2}$. This means that $%
\Delta (\tau )$ can be vanishingly small even for $\tau $ equal to a few
lifetimes, provided that $R_n\gg 1$. In fact, there exists a finite time
interval in which $\Delta (\tau )/|\phi |^2$ is negligible compared to $%
e^{-\tau /2}$, leading to the following exponential buildup law: 
\begin{equation}
|\Psi \left( \tau \right) /\phi |=1-e^{-\tau /2}.  \label{builexp}
\end{equation}

This simple formula reproduces successfully the predicted values of
expression (\ref{Psi1term}). For comparison, we have included in Fig. \ref
{fig2} a plot of the normalized probability density calculated from (\ref
{builexp}). The corresponding curve is indistinguishable from all the other
curves, showing in particular that $\Delta (\tau )$ has an exceedingly small
contribution in the relevant time interval for all of our numerical
examples. We see that formula (\ref{builexp}) does not depend explicitly on
the potential profile parameters nor the resonant state, this explains why
all the numerical examples illustrated in Fig. \ref{fig2} share the same
curve, despite the fact that they correspond to different situations. Note
that in the exponential regime the buildup law becomes identical to the
charging up law of a capacitor in an $RC-$circuit: $Q(\tau )/Q_0=1-e^{-\tau
/\tau _C}$, where $Q_0$ is the asymptotic charge, and $\tau _C=RC$ is the
capacitive time constant. This is relevant, because we find in the
literature capacitor-like models used to describe quantum tunneling
properties in DB structures, such as the charge buildup and its implications
on the speed limit on resonant tunneling devices \cite{capacitor,Luryi}.
According to Eq. (\ref{builexp}), the transient time constant $\tau _0${\em %
\ }of our ``quantum capacitor''{\it \ is always two lifetimes}, and is a
characteristic feature of one-level systems. On the experimental side, it is
worth to mention that measured values of escape times of the order of $%
2\hbar /\Gamma _n$ has been reported by Sakaki {\it et al.} \cite{Sakaki2},
arguing that, at coherence conditions,{\it \ ``the buildup time and the
tunneling escape time are roughly the same'' }\cite{Tsuchiya}. An important
remark is that the condition $R_n\gg 1$ is not so restrictive since it is
satisfied for most of the resonant structures with typical parameters. In
fact we carried out a systematical study (not shown here) and found that for
values of $R_n$ from $10$ onwards this condition is satisfied.

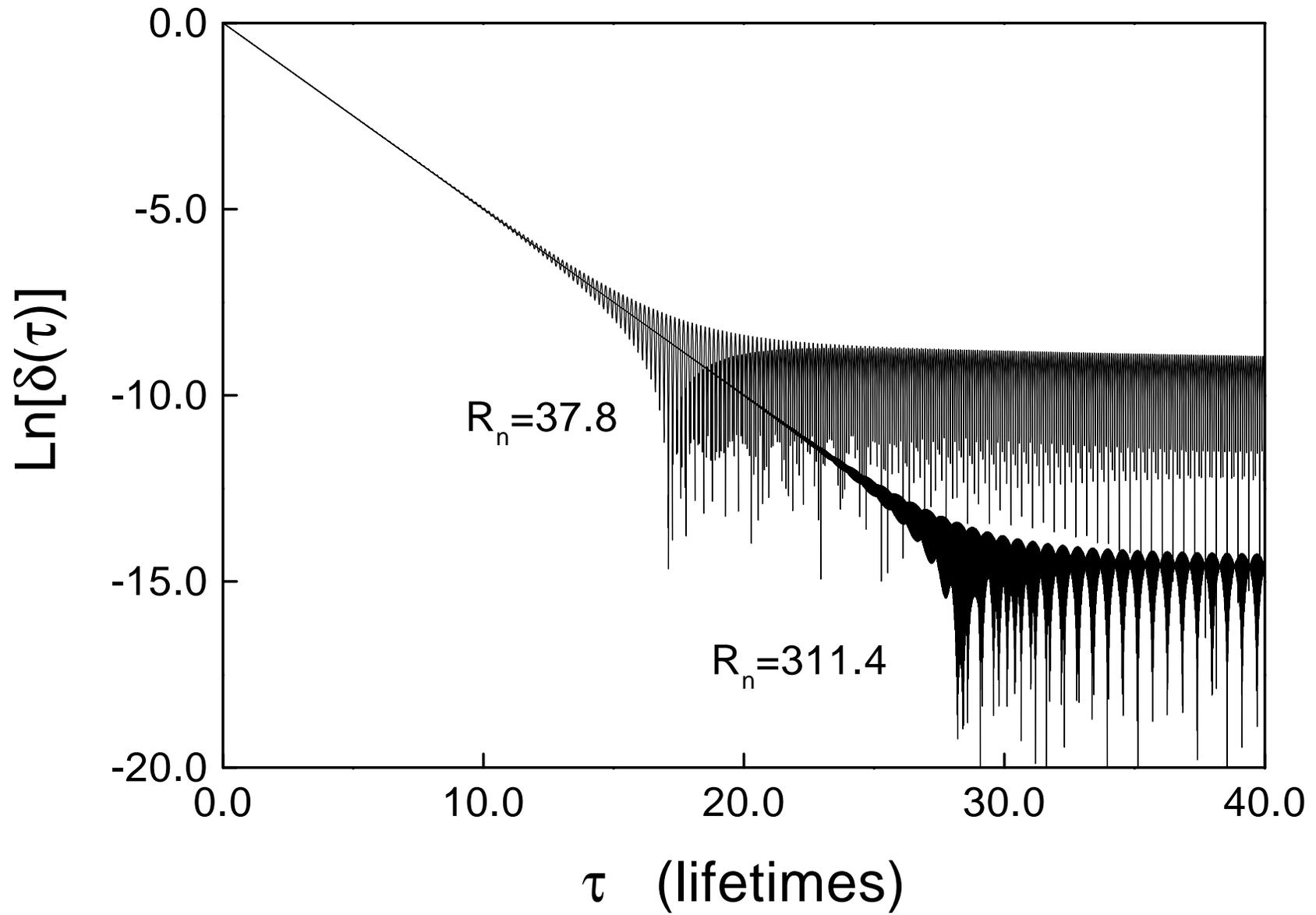
\begin{figure}
\caption{ Exponential and non-exponential contributions to the buildup. We
plot the logarithm of the difference of $\delta (\tau )=|1-|\Psi \left( \tau
\right) /\phi ||$ versus $\tau $, for the states $n=1$ and $n=3$ of the
symmetrical system considered in the text. The corresponding values of the
ratios $R_n\equiv \varepsilon _n/\Gamma _n$ are show in the figure. The
linear behavior corresponds to the exponential regime, and the deviations
from it appear after a certain transition time $\tau _{onset}$. }
\label{fig3}
\end{figure}

In order to show the existence of deviations from the exponential regime, we
shall examine the contributions arising from $\Delta (\tau )$. The explicit
calculation of $\Delta (\tau )$ in terms of $y_{-k}$, $y_{-k_n}$, and $%
y_{-k_n^{*}}$ may result a too involved task, however, for the purpose of
our discussion, it is sufficient to realize that the dominant term is
proportional to an oscillatory function of $\tau $ modulated by the factor $%
\tau ^{-1/2}$. We know that at very long times the exponential term goes to
zero faster than $\tau ^{-1/2}$, {\it i. e.} $e^{-\tau /2}\ll \tau ^{-1/2}$.
Therefore, there must exist a critical time $\tau _{onset}$ at which $%
e^{-\tau /2}$ and $\Delta (\tau )/|\phi |^2$ are comparable. Such a critical
time defines a crossover from the exponential to a non-exponential regime of
the buildup process. In the examples depicted in Fig. \ref{fig2} the
non-exponential contributions to $|\Psi \left( \tau \right) /\phi |^2$ are
overwhelmed and cannot be appreciated due to the scale of the graph.
However, if we plot the logarithm of the difference $\delta (\tau )=|1-|\Psi
\left( \tau \right) /\phi ||$ versus $\tau $, [using Eq. (\ref{Psi1term})]
the transition from the exponential to the non-exponential regime is clearly
appreciated, see Fig. \ref{fig3}. In this figure, the exponential regime can
be identified by the straight line with slope $-1/2$, extending over a
few lifetimes until it reaches the onset of the nonexponential buildup,
$\tau _{onset}$, which depends on $R_n$.

It is interesting to note the similarity of the results depicted in Fig. \ref
{fig3} to the behavior of the {\it survival probability} found in studies of
the phenomenon of quantum decay \cite{NEDtheor,Nico}, which also exhibits
this transition with an oscillatory structure at such crossover. We believe
that this striking resemblance is not a simple coincidence but rather a
manifestation of the existence of a more profound link between both
phenomena. In fact, the survival probability may also be expressed in terms
of the Moshinsky functions \cite{NEDtheor}, which are, in our expressions,
the key ingredients for the time evolution. These findings open up new
questions about the common features in both processes, for example those
about the existence of deviations from the exponential buildup also at early
times, as it occurs in the decay process \cite{NEDshort}. Such analysis
requires the contribution of far away resonances to the transient solution,
and is deferred to future work \cite{unpublish}.

Summarizing: {\it (i)} We have accomplished the first analytic derivation of
the actual buildup law in resonant tunneling structures. It was based on
general properties of the solution of the Schr\"{o}dinger equation, without
any assumptions on the potential profile, except that it is finite and
support well defined resonances. {\it (ii)} We have shown the existence of
both exponential and non-exponential contributions to the buildup process. 
{\it (iii)} The exponential regime is characterized by a transient time
constant whose value is exactly {\it two lifetimes}. {\it (iv)} We have
illustrated that formula (\ref{builexp}) describes very accurately the
exponential buildup for a great variety of situations: it works very well
for different potential profiles, and is valid not only for the ``ground
state'' $\left( n=1\right) $, but also for ``excited states'' $\left(
n>1\right) $.
 
We thank G. Garc\'{\i }a-Calder\'{o}n for useful discussions.

\end{document}